\documentclass{aa}
\usepackage{graphicx}
\usepackage{txfonts}
\usepackage{color}
\usepackage[dvipsnames]{xcolor}
\usepackage{float}
\usepackage{lscape}
\usepackage[colorlinks=true,linkcolor=blue,filecolor=blue,citecolor=blue,urlcolor=blue, draft=False]{hyperref}
\usepackage{natbib}
\usepackage{multirow}

\usepackage{caption}
\DeclareCaptionFormat{cont}{#1#2#3\par}

\newcommand{\citel}[1]{\citeauthor{#1}\,\citeyear{#1}}
\newcommand{\Msun}[1]{\,M$_{\odot}$}
\newcommand{\kmpersec}[1]{\,km.s$^{-1}$}

\usepackage{amstext}

\begin{document} 

   \titlerunning{A catalogue of low-mass X-ray binaries in the Galaxy}
   \authorrunning{F.\,Fortin et al.}

   \title{A catalogue of low-mass X-ray binaries in the Galaxy: from the \emph{INTEGRAL} to the \emph{Gaia} era.\thanks{An online version of the catalogue is publicly available at \url{https://binary-revolution.github.io/LMXBwebcat} and the database in the associated GitHub repository will be continuously updated based on community inputs.}}

   \author{
           F. Fortin\inst1\textsuperscript{,} \inst2 \and
           A. Kalsi \inst3\textsuperscript{,} \inst4\and
           F. Garc\'ia\inst5 \and
           A. Simaz-Bunzel\inst5 \and
           S. Chaty\inst2
          }

   \institute{
   IRAP, CNRS, 9 avenue du Colonel Roche, BP 44346, 31028 Toulouse Cedex 4, France
    \and
   Universit\'e Paris Cit\'e, CNRS, Astroparticule et Cosmologie, F-75013 Paris, France
     \and
     Delhi Technological University, Delhi 110042, India
     \and
     Department of Physics and Astronomy "Galileo Galilei", University of Padua, Italy
     \and
     Instituto Argentino de Radioastronom\'ia (CCT La Plata, CONICET; CICPBA; UNLP), C.C.5, (1894) Villa Elisa, Buenos Aires, Argentina
   }

   \date{received ... ; accepted ...}

 
  \abstract
    {
    Low-mass X-ray binaries (LMXBs) are high-energy sources that require multi-wavelength follow up campaigns to be fully characterized. New transients associated to LMXBs are regularly discovered, and previously known systems are often revisited by astronomers to constrain their intrinsic parameters. All of this information compiled into a catalogue may build up to a useful tool for subsequent studies on LMXBs and their population.
    }
    {
    We provide an update on past LMXB catalogues dating back 16 years and propose to the community a database on Galactic LMXBs with the most complete manually curated set of parameters and their original references. On top of a fixed version accessible through Vizier, we propose to host the catalogue independently on our GitHub collaboration, side-by-side with our previous catalogue on high-mass X-ray binaries. The database will be regularly updated based on new publications and community inputs.
    }
    {
    We build a working base by cross-matching previous LMXB catalogues and supplementing them with lists of hard X-ray sources detected in the past 20 years. We compile information from Simbad on LMXBs as a starting point for a thorough, manual search in the literature to retrieve important parameters that characterize LMXBs. We retrieve newly detected LMXBs and candidates directly from literature searches. Counterparts to these LMXBs are compiled from hard X-rays to infrared and radio domains. Every piece of information presented on the LMXBs is curated and backed by accurate references.
    }
    {
    We present a catalogue of 339 Galactic LMXBs listing their coordinates, companion star spectral type, systemic radial velocity, component masses and compact object nature, the presence of type I X-ray bursts as well as orbital data. Coordinates and identifiers of counterparts at various wavelengths are given, including 140 LMXBs detected in {\it Gaia} DR3.
    }
    {}


   \maketitle
%

\section{Introduction}\label{sect:introduction}

More than 60 years have passed since the identification of the first extrasolar X-ray source \citep{1962PhRvL...9..439G}. This particular source, Scorpius X-1, belongs to the category of low-mass X-ray binaries (LMXB), which are composed of a compact object -- either a black hole (BH) or a neutron star (NS) -- that accretes material from a low-mass companion star ($M$ $\lesssim$ 1\Msun\,). LMXBs are soft X-ray emitters powered by mass transfer happening through Roche lobe overflow (see a general review on accreting binaries in \citel{2022abn..book.....C}). An accretion disk can form around the compact object and can be responsible for the majority of the radiation emitted during periods of high activity from X-rays down to infrared wavelengths. The formation of relativistic jets commonly occurs during specific phases of activity in microquasars (i.e. BH LMXBs). However, these X-ray systems are but a single phase in the whole evolution of binary stars, from their formation up to their endpoint as compact binaries and gravitational wave sources (see a recent review by \citel{2023pbse.book.....T}).

As a follow-up of our previous work on building an updated catalogue of Galactic high-mass X-ray binaries (HMXBs, \citel{2023AA...671A.149F}, now containing 160 sources in HMXBwebcat\footnote{\url{https://binary-revolution.github.io/HMXBwebcat}}), we present here our latest catalogue of 339 LMXBs in the Milky Way. Information available on LMXBs in the literature suffers from the same caveat as HMXBs: as these sources are inherently difficult to observe and require multi-wavelength and/or time series campaigns to be properly constrained, most of the discoveries on LMXBs end up spread over several years and can come from many different teams of researchers across the world. This is why the catalogue presented in this paper is centred on two important concepts: firstly, for each parameter listed, we provide the original reference where it was derived : this results in a compilation of more than 600 unique, curated references. Secondly, in parallel to being hosted in Vizier in a fixed version, we host it independently on a dedicated website, LMXBwebcat, on which the database is able to receive updates through our GitHub collaboration\footnote{\url{https://github.com/Binary-rEvolution/LMXBwebcat}} as new observations are made, and new or updated parameters of already known sources become available. Hence this catalogue is open-source and we invite the community to participate in adding their discoveries and/or updates on LMXBs, or correct any mistake/omission they may identify in our catalogue.

Since the late 90s/early 2000s, X-ray observatories such as {\it INTEGRAL}\footnote{INTernational Gamma-Ray Astrophysics Laboratory}, {\it XMM-Newton}\footnote{X-ray Multi-Mirror Mission}, {\it Chandra}, {\it Swift} and {\it MAXI}\footnote{Monitor of All-sky X-ray Image} have participated in discovering new LMXBs and gaining in-depth knowledge on their intrinsic parameters such as orbital data, companion mass, or the exact nature of the accreting compact object. All of this information would benefit from being compiled together to facilitate further studies on LMXBs. But the need for such a catalogue is not only motivated by the observational progresses of the past 16 years; it is also the upcoming facilities dedicated to the high energy and transient sky, such as {\it eROSITA}, LSST\footnote{Large Synoptic Survey Telescope} and {\it SVOM}\footnote{Space Variable Objects Monitor}, that warrant keeping a constantly up-to-date catalogue in anticipation of the amount of new data that will be available on X-ray binaries. In turn, this will also allow us to better predict the population of gravitational wave sources that LISA\footnote{Laser Interferometer Space Antenna} will be able to detect (see the white paper \citel{2023LRR....26....2A}).

The first instances of LMXB catalogues date back to the 1980s (\citel{1983ARAA..21...13B}; \citel{1983adsx.conf..192V}) and counted roughly 33 X-ray systems. Subsequent catalogues were made following the advancements of space-based X-ray observatories, namely the ones presented in \cite{1995xrbi.nasa..536V}, \cite{2001AA...368.1021L}, and finally the latest and most used to date catalogue of 187 Galactic LMXBs by \cite{2007AA...469..807L}.

In this paper, we first describe the method we followed to build the catalogue (Section\,\ref{sect:catalogue}, which follows a similar structure to what we presented in our HMXB catalogue \citel{2023AA...671A.149F}), namely for the recovery of parameters and numerous counterparts to each LMXB. We discuss in Section\,\ref{sect:byproducts} the catalogue itself and how it compares to previous iterations and other similar catalogues. We also discuss the potential uses of our catalogue, and conclude on this work in Section\,\ref{sect:conclusion}.

\section{Building the catalogue}\label{sect:catalogue}

In this section we provide a step-by-step description of the tasks we performed to build the catalogue. We have adopted a very similar approach presented in the twin catalogue dedicated to HMXBs \citep{2023AA...671A.149F}, such as using pre-existing catalogues of high energy sources as a starting base, relying on the services of the Centre de Données Astronomiques de Strasbourg (CDS) such as Simbad and Vizier, and of course minute manual search through references using NASA's Astrophysics Data System (ADS) abstract search engine.

\subsection{Reference catalogues and building a working base}\label{subsect:base}
More than 16 years after publication, the LMXB catalogue by \cite{2007AA...469..807L} is still the most used reference; at that time, the authors listed 187 LMXBs and candidates in the Milky Way. We supplemented this list with the catalogue of {\it INTEGRAL} detections \citep{2016ApJS..223...15B} which contains 939 hard X-ray sources of various nature. We only kept the sources labelled LMXB or unidentified, thus adding 345 sources. We identified sources common to both \cite{2007AA...469..807L} and \cite{2016ApJS..223...15B} using positional and identifier cross-match performed with TOPCAT \citep{2005ASPC..347...29T}. We then queried Simbad for all the sources containing the type (or subtype) "LXB"; this returns a list of 674 sources, most of which are extragalactic. These are usually grouped in small clusters around their host galaxy, hence we used the same spatial filtering method as presented in \cite{2023AA...671A.149F}, where we check each source for the presence of close neighbours. However, contrary to HMXBs, LMXBs are known to be present in globular clusters, and these structures may hold up to $\sim$10 LMXBs very close together (as  it is the case in NGC 6440 for instance). Hence, we used the list of globular clusters available in \cite{2021MNRAS.505.5978V} to check the likelihood of a group of LMXBs to be part of a Galactic globular cluster, before filtering them out if they have neighbours closer than 10\arcmin. This way, we retrieved 251 Galactic LMXB candidates from Simbad.

Eventually, after cross-matching the Simbad LXBs with the sample of high energy sources from \cite{2007AA...469..807L} and \cite{2016ApJS..223...15B}, we obtain a working base of 573 sources. We note that a significant amount (216) were listed in the {\it INTEGRAL} catalogue as unidentified, and most of them will be discarded through a thorough manual process of checking new information published on these sources in the literature.

\subsection{Retrieving binary parameters and new LMXBs}\label{subsect:parameters}

While pieces of information are available on the LMXBs and candidates compiled so far (mainly from the data catalogued in \citel{2007AA...469..807L}, \citel{2011yCat....102018R} and Simbad), our goal for this catalogue is to perform a thorough investigation on each of these sources to, firstly, identify which are interesting LMXB candidates, and secondly, retrieve useful parameters as well as the original studies that derived them (this includes double checking references already compiled by former catalogues).  We find that the later point is especially important, as some parameters listed might have been derived using different methods for each source; this information should be readily available for authors of further studies that revisit these sources. Thus, during the individual search for information on each of these sources, we only use already available data as a starting point for the manual search, and always re-check already available references. We compile new and updated data whenever they are available, confirm the already measured parameters by retrieving the original studies, and search for older publications to ensure we have compiled every pieces of information possible on each binary. Hence, none of the parameters or references we list in the catalogue come from an automated, unsupervised query on Simbad. We did use the resources in works that already compiled some useful data on LMXBs, such as spectral types, orbital periods and masses in \cite{2008NewAR..51..860Y}; a list of type I X-ray bursts seen be RXTE in \cite{2008ApJS..179..360G} and its extension to {\it BeppoSAX} and {\it INTEGRAL} MINBAR \citep{2020ApJS..249...32G}; orbital periods in \cite{2022PASJ...74..974A}; the catalogue of stellar-mass black holes in X-ray binaries BlackCAT \citep{2016AA...587A..61C}; or the latest catalogue of Ultra-Compact X-ray Binaries UltraCompCAT (\citel{2023arXiv230507691A}, which is also a dynamic database\footnote{Available at \url{https://research.iac.es/proyecto/compactos/UltraCompCAT/}}).

Because LMXBs can be highly variable sources, we settled on a list of parameters to look for we reckon are representative of the systems at all times. These are the spectral type of the companion star, the nature of the accreting compact object, masses of the binary components, the orbital parameters, the radial velocity of the center of mass, the presence of a spin period and its time derivative, and the detection of type-I X-ray bursts. We try not to log any "assumed" values for any parameters --such as when NSs are assumed to be 1.4\,\Msun\, in orbital solutions or when the LMXBs are being located at 8\,kpc by default if they are near the Galactic center-- unless they are motivated or indirectly constrained by other measurements.

Lastly, we performed "blind" searches within the literature for newly detected LMXBs and candidates in papers ranging from 2016 to 2023 that may not appear in either catalogues of high energy sources we used or not yet be logged as LMXBs in Simbad. This was mainly done through queries of keywords such as "binary", "transient", "X-rays" in the NASA ADS abstract service. Most new candidates discovered this way come from Astronomer's Telegrams.

We note that at this point, we have made the data in the catalogue as reliable as possible; this kind of curated data-mining is however bound to contain small errors or oversights. This is why the catalogue is hosted on an independent website and, as our previous catalogue already did \citep{2023AA...671A.149F}, will receive gradual updates and new releases as new information is either found by ourselves or brought forward by the community.

\subsection{Finding an unambiguous chain of counterparts}\label{subsect:counterparts}
As we previously argued in \cite{2023AA...671A.149F}, we believe that part of what consists of a secure identification as LMXB or candidate LMXB is having an unambiguous list of counterparts, ideally from hard X-ray to infrared. Having this positional information is also a necessary tool for astronomers to prepare new observations in follow-up campaigns. The automated search for counterparts to each LMXB is not trivial, but still facilitated by positional data already present in Simbad, and also by the stored identifiers which also contain information about the coordinates of the source and their accuracy. Our goal here is to perform a cone search on all the catalogues listed in Table\,\ref{tab:counterparts} in order to retrieve a proper chain of counterparts from the high energies down to the infrared wavelengths. For this, we need an initial set of coordinates to query around in these catalogues.

These starting coordinates are either the ones listed in Simbad, or the ones that we parsed from the list of identifiers available in Simbad, depending on whichever is most accurate. For instance, the candidate LMXB IGR J17480$-$2446 has coordinates available in Simbad, but without any information on their accuracy. It does have another identifier, CXOGlb J174804.8$-$244648, which we can parse into a set of coordinates with an accuracy below the arcsecond scale as it comes from {\it Chandra} observations. This is a quick method that is capable of automatically retrieving accurate positional data from Simbad, that may not necessarily be available in catalogues but which was for instance derived in Astronomer's Telegrams. In this particular case, this allowed us to instantly find the {\it Chandra} counterpart of IGR J17480$-$2446 (2CXO J174804.8$-$244649) with a cone search in the Chandra CSC 2 database \citep{2019yCat.9057....0E}.

This positional cross-match is still prone to false positives, as we have to purposefully look within the catalogues on a sufficiently large area so that we do not miss any potential counterparts. This is why we also perform a recursive search within the produced list of counterparts, from the poorest to the most accurate catalogues ({\it XMM-Newton}, {\it Chandra}, 2MASS, {\it Gaia}). We note that for LMXBs closely grouped within globular clusters, this method is sometimes unable to automatically separate different chains of counterparts, especially for hard X-rays for which the astrometric precision is not accurate enough to separate closely grouped LMXBs in clusters. Thus, for LMXBs found in clusters, we manually checked the consistency of their counterparts to ensure each have a unique, precise localisation.

\begin{table}[h]
    \centering
    \caption{List of queried catalogues for the counterpart search.}
    \label{tab:counterparts}
    \begin{tabular}{lll}
        \hline\hline\\[-2ex]
        Catalogue & Reference & Radius \\
        \hline\\[-2ex]
        \textit{HEAO 1} & \cite{1984ApJS...56..507W} & 20\arcmin \\
        \textit{Uhuru 4}  & \cite{1978ApJS...38..357F} & 20\arcmin\\
        \textit{Ariel V 3} &  \cite{1981MNRAS.197..865W} & 20\arcmin \\
        \textit{INTEGRAL}  & \cite{2016ApJS..223...15B} & 20\arcmin \\
        \textit{Fermi}  & \cite{2022ApJS..260...53A} & 20\arcmin \\
        \textit{BeppoSAX}  & \cite{2011ApJS..195....9C} & 6\arcmin  \\
        \textit{Einstein 2E} &  \cite{1990EObsC...2.....H} & 4\arcmin \\
        \textit{ROSAT}  & \cite{2000yCat.9031....0W} & 35\arcsec \\
        \textit{Swift} 2SXPS  & \cite{2020ApJS..247...54E} & 8\arcsec \\
        4XMM DR11 &  \cite{2020AA...641A.136W} & 4\arcsec \\
        \textit{Chandra} CSC 2 &  \cite{2019yCat.9057....0E} & 3\arcsec \\
        2MASS &  \cite{2003yCat.2246....0C} & 120\,mas \\
        \textit{Gaia} DR3 &  \cite{2022yCat.1355....0G} & 20\,mas \\
        \hline
    \end{tabular}
\end{table}

\subsection{Contents of the catalogue}\label{subsect:contents}

The general characteristics of Galactic LMXBs are presented in Table\,\ref{tab:cat:general}. The full catalogue provides their Simbad identifier ("Main ID" field from Simbad), as well as a short list of the most used IDs we found in the literature. To build this list, we queried each known IDs from Simbad in ADS and retrieved how many papers used them in their title, abstract or full text. We rank them from most popular to least popular, and thus provide identifiers that should reflect the community's preferred naming convention. In this paper, we list the LMXBs under their most popular identifier, except for a few sources for which no popular ID can be recovered in ADS; in this case, we use their Simbad identifier. The online catalogue provides the full list of identifiers known by Simbad, to facilitate queries in LMXBwebcat. The "Compact" column provides information on the nature of the accreting compact object, and "Spectype" refers to the spectral type of the donor star. When available, we list the distance inferred from {\it Gaia} EDR3 parallaxes by \cite{2021AJ....161..147B}. We also have cross-matched our LMXBs with the list of globular clusters from \cite{2021MNRAS.505.5957B}, which contains information about their distances inferred from various means, including {\it Gaia} parallaxes. Hence we provide, in case of a spatial association with a globular cluster, the name of the cluster and the inferred distance in the "Other distance" column. In the text, we only keep the largest error bound for {\it Gaia} distances and values are voluntarily rounded for the sake of readability; the unaltered numbers are available in the electronic versions of the catalogue. When there is no cluster association, the "Other distance" column may also contain distance estimates from a number of other methods, which are available in the given references. Lastly, we indicate when one or several type I X-ray burst were observed by providing a reference.

In Table\,\ref{tab:cat:orbital} are presented the orbital parameters: orbital period, eccentricity and semi-amplitude of the donor's radial velocity. We also give details on the mass of each binary component --"Mx" for the compact object, "Mo" for the companion star-- as well as information on the presence of a pulse period and its time derivative in the case the compact object is a NS. Again, values are rounded for readability and are available in full in the electronic version of the catalogue.

The electronic version also provides extra information on the counterparts for each binary, namely their numerous identifiers, coordinates and corresponding astrometric precision. The right ascension and declination are given in J2000, and astrometric uncertainties are the 90\% positional errors retrieved from the queried catalogues. The full contents of the LMXB catalogue can be queried on Vizier, or viewed and queried from the dedicated website LMXBwebcat\footnote{\url{https://binary-revolution.github.io/LMXBwebcat/catalog.html}}. The latter database will be updated regularly and new versions of the catalogue will be published on the website; every changes will be logged to track the evolution of each version, which will remain available alongside the latest upload.

We advise the readers to remain critical of all the parameters we list in this catalogue, and to check the references before using them in their own work. To ease the process, we use a flag system to quickly inform the reader about the reliability of each listed parameter. The flags appear as numbers in dedicated columns in the catalogue database; for the sake of readability, we use daggers in the text:
\begin{itemize}

\item 0 / (nothing): directly measured parameter. This flag is not displayed in Tables\,\ref{tab:cat:general} and \ref{tab:cat:orbital} nor in LMXBwebcat. This includes companion spectral types inferred through spectroscopy, full orbital solutions derived from radial velocity followups, orbital periods from pulse timing or eclipses, dynamical masses, NSs confirmed by the presence of type I bursts or a spin period, or compact object type identified through the measure of the dynamical mass. This also includes distances of globular clusters hosting LMXBs derived via the Gaia parallaxes of the cluster members.
\item 1 / ($\dagger$): caution. The parameter is either indirectly measured or needs better constraints. This includes tentative orbital periods, rough estimations of spectral types from photometry, compact object masses and distances from spectral fitting, radial velocity semi-amplitudes from empirical relations, or compact object types identified through comparison of X-ray spectra. It also includes all parameters for which we only have upper or lower limits on a direct measurement (i.e. most eccentricity measurements).
\item 2 / ($\dagger\dagger$): warning. The parameter is assumed in the scope of a model and likely needs proper observational constraints. This also includes indirectly derived parameters with lower or upper limits.

\end{itemize}

For instance, out of the 75 BH LMXBs we list, only 19 are flagged as reliable BH ("0") since direct evidence of the presence of a BH is hard to get, apart from deriving a dynamical mass greater than $\sim$3\,M$_{\odot}$. Most BHs are thus flagged "1" (N=52) and only a few flagged "2" (N=4) since this identification mostly comes from their X-ray spectrum resembling the one of typical BH LMXBs. The presence of a NS is however much easier to prove through the observation of type I X-ray bursts or the detection of a spin period; hence 146 out of 176 NSs in LMXBs are flagged as being reliable. These flags can be used to quickly identify which LMXBs are worth revisiting to give better constraints on their set of parameters.

\subsection{Gaia candidate counterparts}\label{subsect:gaia}
The only parameter flag that slightly differs from the others is the one attributed to the \textit{Gaia} distances in Table\,\ref{tab:cat:general}.
Because LMXBs are generally very faint in optical/nIR, and because there are so many late-type stars in the field, finding the true optical/nIR counterpart to an LMXB can be challenging even in the case sub-arcsecond X-ray localisation is available, as interlopers can be present along with the true counterpart within the X-ray error circles. Hence we propose here candidate \textit{Gaia} counterparts to LMXBs based on astrometric cross-match as well as historical associations in the few cases where both deep and high-resolution optical/nIR imaging has been performed.

The Gaia distance flag does not refer to the reliability of the distance estimation; instead it represents how the \textit{Gaia} counterpart was associated to the LMXB. \textit{Gaia} counterparts that we found compatible with either an \textit{XMM-Newton} or \textit{Chandra} detection (e.g. a sub-arcsecond scale association), and that do not have any other neighbour in the \textit{Gaia} catalogue closer than 1\arcsec, are deemed secure and as such we do not flag them. \textit{Gaia} counterparts that were found only based on a \textit{Swift} detection (e.g. an arc-second scale association) are flagged as uncertain ("1"/$\dagger$). The \textit{Gaia} counterpart flagged as "2"/$\dagger\dagger$ are not necessarily unreliable, but are not based on an association with an accurate soft X-ray detection. Instead, they come from historical association of an optical counterpart. For instance, Sco~X-1 has a Gaia counterpart but no soft X-ray position available in the \textit{Swift}, \textit{Chandra} or \textit{XMM} catalogues because of its tremendous X-ray brightness.

\begin{figure}
    \centering
    \includegraphics[width=\columnwidth]{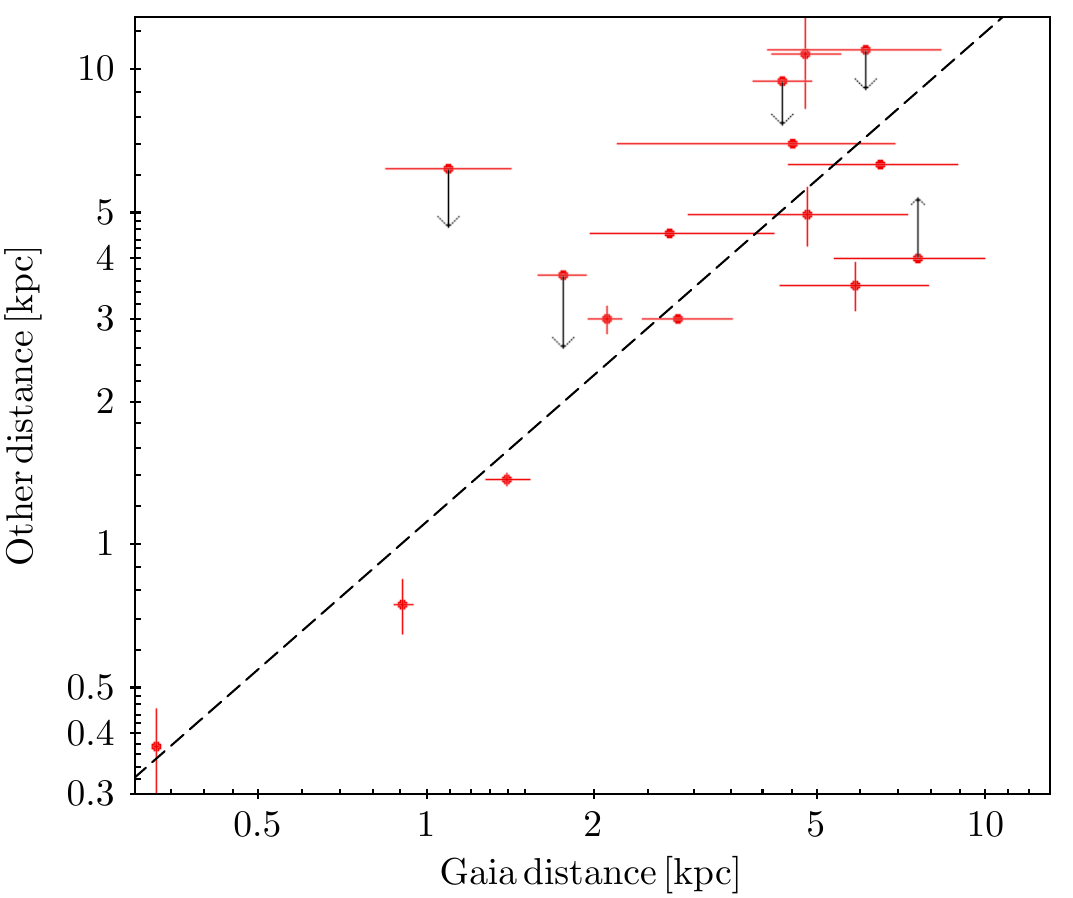}
    \caption{Comparison between the distances to LMXBs from the literature vs. the distances of the proposed \textit{Gaia} counterparts inferred from their parallax. Only LMXBs that have both information and are not located within a cluster are plotted (N=16); black arrows indicate lower/upper limits. The dashed black line is the best linear fit, excluding sources with only lower/upper limits (N=11).}
    \label{fig:distances}
\end{figure}

These \textit{Gaia} candidate counterparts can be subject to change in further iterations of the database as new X-ray positional constraints and/or deeper optical/nIR imaging becomes available on LMXBs. There are ways to further determine the level of confidence in the \textit{Gaia} counterparts we found, such as comparing their optical magnitudes to previously published survey data. Because LMXBs can be highly variable, we chose to perform a check between the distances inferred by \textit{Gaia} parallax and the distances inferred by other means ("other distance", such as spectral energy distribution fitting or photospheric radius expansion in bursting sources). After removing the LMXBs that lie within clusters (since all of their distance information come from \textit{Gaia} data), we plot in Figure\,\ref{fig:distances} the Gaia distances vs. the other distances from the literature. A linear fitting excluding the data points with lower/upper limits returns a proportional coefficient of 1.03 and a systematic of 50\,pc, which is satisfactory since we are comparing two completely independent sets of measurements. We note that this sub-sample is representative of the whole catalogue concerning the \textit{Gaia} flags: we count four flag-0, a single flag-1 and six flag-2 \textit{Gaia} counterparts in the fitted sample. This may indicate that our flagging system is a bit too strict, with the caveat of a low number of sources to work with.

\section{Results, discussions, and byproducts}\label{sect:byproducts}
\subsection{Catalogue statistics and uses}\label{subsect:stats}

\begin{figure}[h]
    \centering
    \includegraphics[width=\columnwidth]{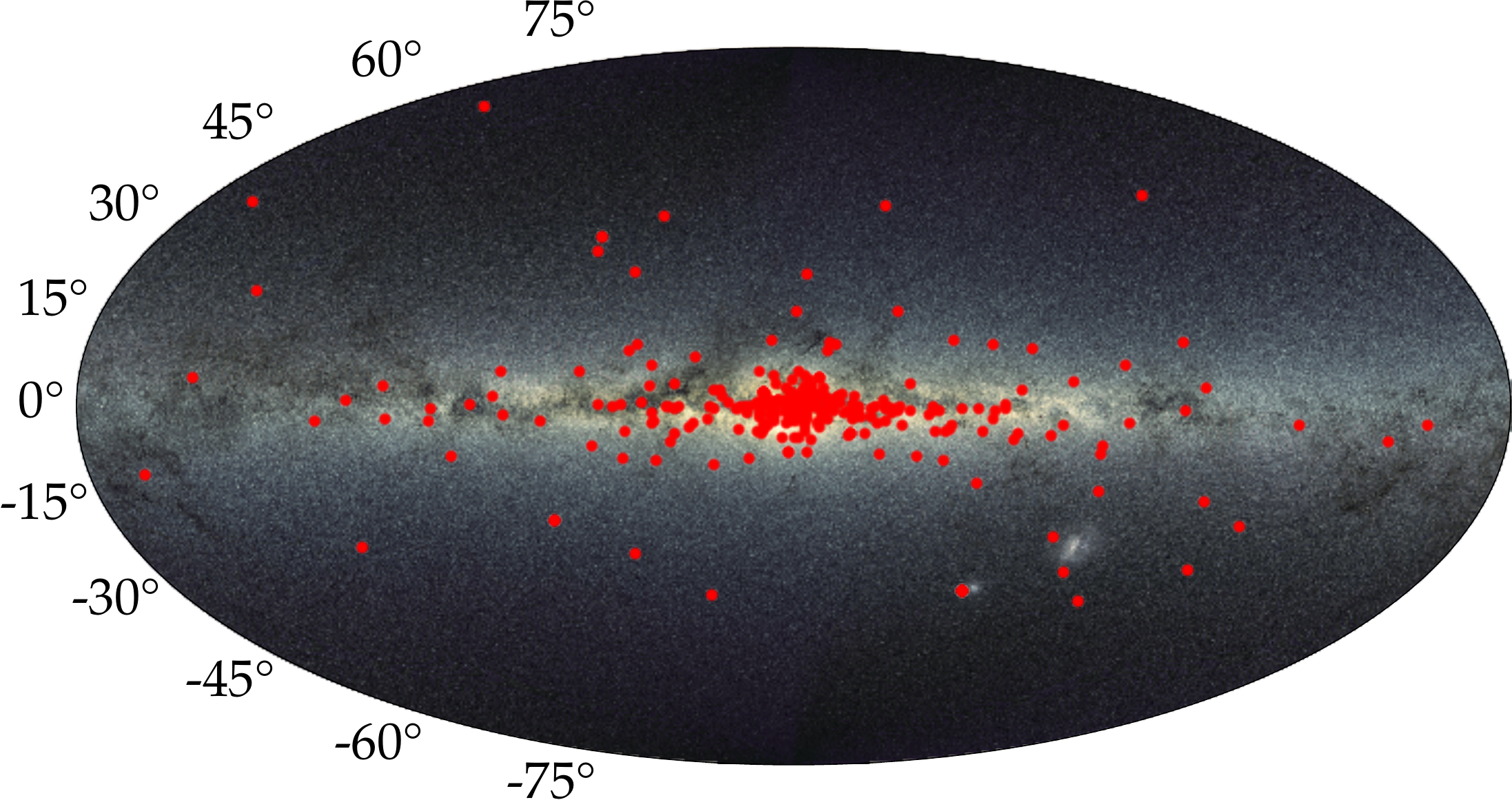}
    \caption{Edge-on view of the 339 LMXBs in the Galaxy. Galactic latitudes are indicated in degrees. \textit{Background image credits: ESA/Gaia/DPAC.}}
    \label{fig:edgeon}
\end{figure}

We present a list of 339 Galactic LMXBs and candidates in this new catalogue, as represented in Figure\,\ref{fig:edgeon}. We note that contrary to the previous catalogue of \cite{2007AA...469..807L}, we do include quiescent LMXB candidates (qLMXBs hereafter); 25 are present in the current version of our catalogue. Hence, after subtracting these qLMXBs to compare our total sample to \cite{2007AA...469..807L}, we provide a 67\% increase in the total number of LMXBs and candidates in the Milky Way. We have at least a tentative identification of compact object type for more than 250 LMXBs, with a predominance of NSs (70\%) compared to BHs (30\%). We compile 150 orbital periods and 69 spin periods, which are respectively a 200\% and 150\% increase compared to the 2007 catalogue, showing the tremendous impact of new detections and follow-up studies on LMXBs. Out of the 176 NS LMXBs, 112 are known to be X-ray bursters.

\begin{figure}
    \centering
    \includegraphics[width=\columnwidth]{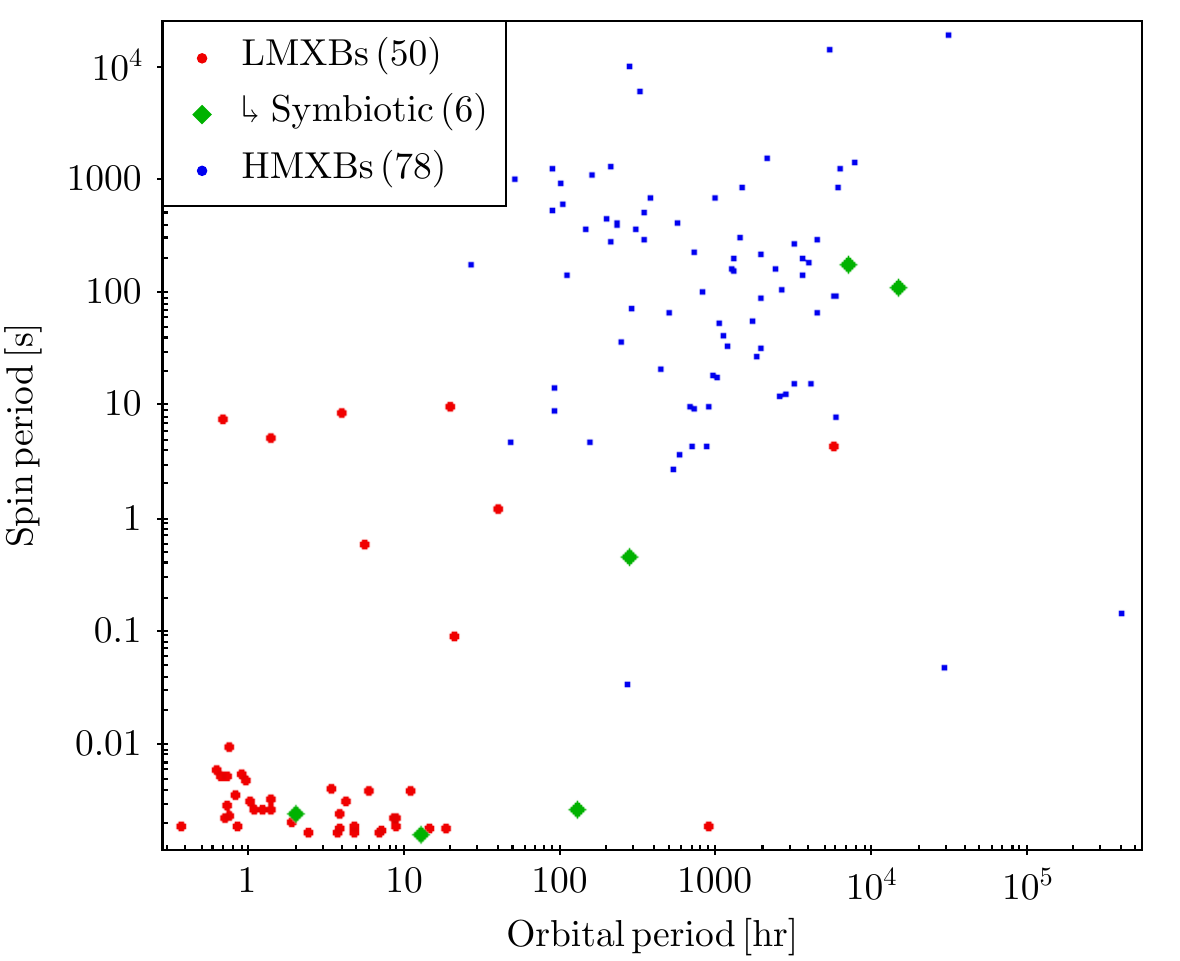}
    \caption{Corbet diagram of the 50 NS LMXBs in the current catalogue, alongside the 78 NS HMXBs now available in \cite{2023AA...671A.149F} which have both orbital and spin period information (all flags included). Symbiotic LMXBs are indicated in green rhombuses.}
    \label{fig:corbet}
\end{figure}

In Figure\,\ref{fig:corbet} we present the full Corbet diagram of NS XRBs in the Galaxy, compiling the data of the present catalogue for NS LMXBs with both orbital period and spin period derived (N=50) alongside the same data for HMXBs available in \cite{2023AA...671A.149F}. In this sample, the majority of the NSs in LMXB have spin periods lower than 10\,ms (N=39), a region exclusively dominated by LMXBs contrary to NSs in HMXBs that seldom have spin periods lower than 100\,ms. LMXBs and HMXBs are generally well-separated in the Corbet diagram, although some symbiotic LMXBs (where the donor is an evolved, giant star) do overlap with HMXBs as consequence of larger orbital separation and lower efficiency at transferring angular momentum.

\begin{figure}
    \centering
    \includegraphics[width=\columnwidth]{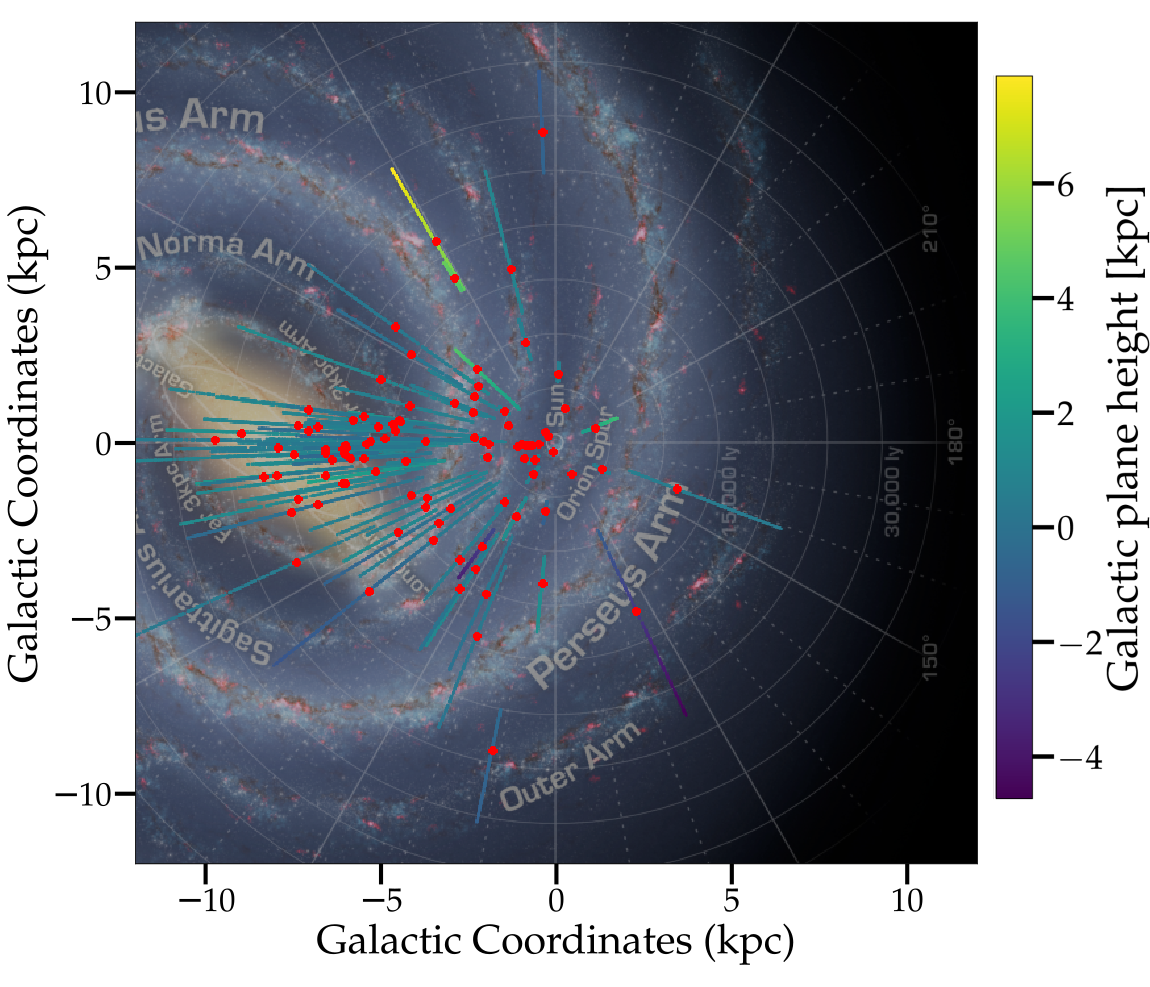}
    \caption{Face-on view of the 98 Galactic LMXBs with \textit{Gaia} parallaxes. Bars indicate the 68\%\,confidence interval in distance. \textit{Background image credits: NASA/JPL-Caltech/R. Hurt (SSC/Caltech)}}
    \label{fig:faceon}
\end{figure}

In our previous catalogue dedicated to HMXBs, almost 3/4 of the HMXBs were detected by {\it Gaia} and had sufficient astrometric quality to invert distance from parallax; in this LMXB catalogue we note that only about 40\% of the sources (140) have a {\it Gaia} DR3 counterpart and only 98 of them have distance information available in \cite{2021AJ....161..147B}. This is explained by the fact that optical counterparts of LMXBs are fainter on average, and also because these are older astrophysical objects that had time to migrate out of the Galactic Plane towards the bulge of the Milky Way, which is too far away for any source to have a reliable {\it Gaia} parallax even when a counterpart is detected. This still allows us to localise $\sim$30\% of our LMXBs within the Galaxy thanks to \textit{Gaia} as shown in Figure\,\ref{fig:faceon}. If we combine to the {\it Gaia} distances the ones found in the literature and the distances to globular clusters, we reach the number of 208 LMXBs with a distance estimation. We note that regardless of the method used for distance estimation (parallax, globular cluster association or X-ray modelling), we advise the users to remain cautious of the determined distance values when considering individual systems as they may be subject to various biases, such as poor parallax quality, wrong association within a globular cluster, foreground star, etc. When considering the whole sample, these biases should at least partly cancel out, but there are bound to be outliers.

\begin{figure}
    \centering
    \includegraphics[width=\columnwidth]{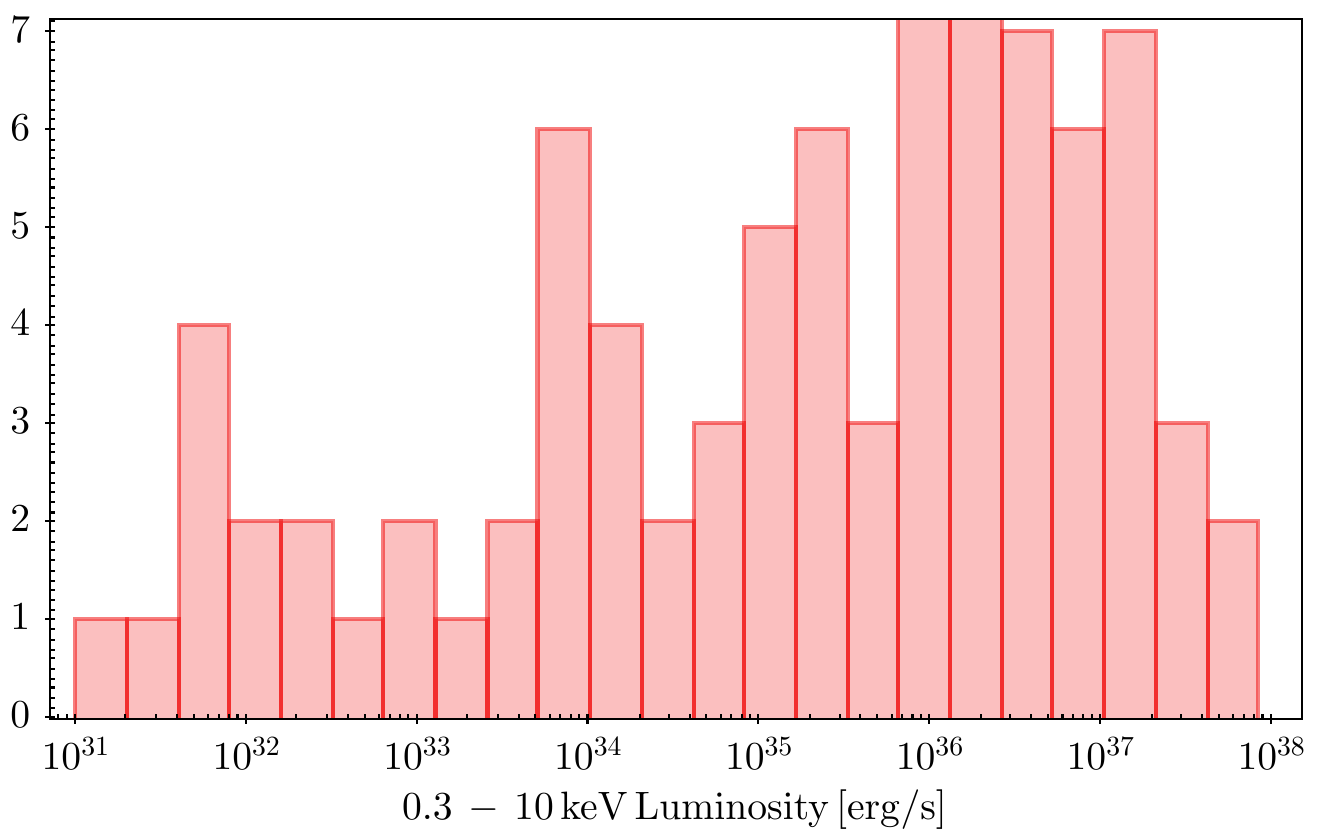}
    \caption{Distribution of the soft X-ray luminosities of Galactic LMXBs seen by \textit{Swift} and with a distance determination (N=89).}
    \label{fig:xlum}
\end{figure}

To show the potential of the data we aggregated in this catalogue, we constructed the distribution of X-ray luminosities of the Galactic LMXBs using the 2SXPS \textit{Swift} database \citep{2020ApJS..247...54E}. We combined the Unabsorbed 0.3--10\,keV Apec Flux (labelled "FAU0" in 2SXPS) with either the {\it Gaia} distances or, when not available, the "Other distance" determination in the LMXB catalogue. In the case the LMXBs are associated with globular clusters, we use the distance to the cluster. This luminosity distribution, shown in Figure\,\ref{fig:xlum}, is the first step to get an estimation of the total X-ray budget of LMXBs in the Milky Way, and is only presented here as an example of how this catalogue can be used for. To go further, users may want to consider other methods of flux measurements from the 2SXPS catalogue, combine fluxes from other observatories such as \textit{XMM-Newton} \citep{2020AA...641A.136W} or \textit{Chandra} \citep{2019yCat.9057....0E} and of course discuss the impact of the behaviour of LMXBs from quiescence to outburst on the average flux values available in the aforementioned catalogues.

\subsection{Byproducts}\label{subsect:byproducts}
A noteworthy byproduct of our work is the retrieval of information about {\it INTEGRAL} sources marked as unidentified in \cite{2016ApJS..223...15B}. We kept five of them as candidate LMXBs, which appear in our catalogue; we also provide in Table\,\ref{tab:IGR} a separate list for the 44 {\it INTEGRAL} sources for which we found an identification in the literature that is not present in Simbad; most are background Active Galactic Nuclei (AGNs) or Cataclysmic Variables (CVs). The purpose of this table is to provide easily digestible data for the CDS to update references in Simbad.

\subsection{Discussion}

After the publication of our previous HMXB catalogue and during the writing of this paper, a similar LMXB catalogue was released by \cite{2023AA...675A.199A}, where the authors list 349 Galactic LMXBs.
We have similar approaches, as the authors also propose to update their database based on community inputs. We did find however notable differences: first and foremost,
we do not cite previous catalogues as references for the parameters in our catalogue; instead, we look for the original studies cited in \cite{2007AA...469..807L} or \cite{2003AA...404..301R} and update them when necessary. This results in a compilation of 570 unique references for the parameters we compile in the present catalogue.
Secondly, we carefully review any reference provided by Simbad, as we have found many instances where they do not point to the original study deriving the cited parameter.
For example, several companion spectral types are given in \cite{2008NewAR..51..860Y}, which is a paper discussing evolutionary models for BH LMXBs that happens to list these spectral types (and some other parameters such as orbital periods or mass ratios) coming from already published data.
Thirdly, we are slightly more strict concerning the inclusion of LMXB candidates,
as we do not list X-ray sources for which there is no particular indication of them being at least candidate LMXBs. For this reason, we do not include in our catalogue: AX J1824.5$-$2451, [BSP2003] 24, Swift J174038.1$-$273712, CXOGC J174538.0$-$290022, XMM J174544$-$2913.0, SWIFT J174553.7$-$290347, CXOGC J174622.2$-$290634, Swift J175233.9$-$290952 and NGC 6752 CX19.
Thanks to our minute investigation of each LMXB candidates, we were also able to exclude doubloons where one or several counterparts of the same source were listed as independent LMXBs. This was the case for KS 1741$-$293, IGR J17353$-$3539, TYC 6824$-$713$-$1, [ZGV2011] 9 and 3U 1728$-$16.
Lastly, we provide corrected identifications for some sources that may have been wrongly reported to be LMXBs, such as the HMXB Cir X-1 \citep{2007MNRAS.374..999J}, the cataclysmic variable OGLE BLG-ELL-12042 \citep{2021MNRAS.502...48G}, and some which are more likely to not be LMXBs ([PLV2002] CX10 is a candidate CV in \citel{2002ApJ...573..184P}).

It is of prime importance for a catalogue to carefully list references. Firstly, this provides at least some insurance to the users that the data was manually curated as, for now, there are no machine learning algorithm that outperform manual data mining in the literature as we present here. Secondly, we also find that it is important to acknowledge all the observational work done on LMXBs and X-ray binaries in general since, as the current catalogue shows, there are still many missing parameters on X-ray binaries that will only be derived thanks to regular proposals for follow up observations by teams of astronomers. Lastly, the main motivator for such catalogues is to untangle the huge amount of information spread within the literature; citing previous catalogues as a source for data goes against this endeavour of making data more accessible to the community.

\section{Conclusion}\label{sect:conclusion}

The continuous endeavour of catching X-ray transients and performing multi-wavelength follow ups has brought the discovery of many new LMXBs and helped characterize their properties tremendously in these past 16 years. To bring together all this information, we present a new catalogue of 339 Galactic LMXBs that constitutes a tool for further studies on either individual systems or their whole population. Being hosted on an independent website, LMXBwebcat, this catalogue is also ready to receive updates/corrections based on new publications or input from the community. Compared to the 187 LMXBs listed in \cite{2007AA...469..807L}, the current number of known Galactic LMXBs has almost doubled, and given the upcoming observational landscape dedicated to high energies and/or transient sky astronomy ({\it eROSITA}, LSST or {\it SVOM} for instance), these numbers are likely to keep growing further.

Thus, it is essential to continue monitoring new discoveries on X-ray binaries so that we are able to better grasp their properties as a population, their role in the Galactic ecology and how, at the endpoint of their evolution, they may become gravitational wave sources detectable by LISA \citep{2018PhRvL.121m1105T}. We would also like to note that optical/infrared follow ups are still much needed to complete the set of observables (such as spectral types or orbital solutions from radial velocities) available on LMXBs. We hope this catalogue can also be a tool to identify which parameter needs constraints and to facilitate the identification of ideal targets for astronomers to revisit.

\begin{acknowledgements}
We thank the anonymous Referee for their meticulous review of both the manuscript and the database, which allowed us to greatly improve them.
The authors were supported by the LabEx UnivEarthS: Interface project I10 "Binary rEvolution: from binary evolution towards merging of compact objects".
SC is grateful to the CNES (Centre National d'\'Etudes Spatiales) for the funding of MINE (Multi-wavelength \textit{INTEGRAL} Network).
FG is a CONICET researcher and acknowledges support from PIP 0113 and PIBAA 1275 (CONICET).
This work made use of NASA's Astrophysics Data System (ADS) web services, and of the services associated to the Centre de Données Astronomiques de Strasbourg (CDS) Simbad and Vizier.
This work has made use of data from the European Space Agency (ESA) mission
{\it Gaia} (\url{https://www.cosmos.esa.int/gaia}), processed by the {\it Gaia}
Data Processing and Analysis Consortium (DPAC,
\url{https://www.cosmos.esa.int/web/gaia/dpac/consortium}). Funding for the DPAC
has been provided by national institutions, in particular, the institutions
participating in the {\it Gaia} Multilateral Agreement.

{\em Software:} Topcat \citep{2005ASPC..347...29T}, {\sc matplotlib} \citep{hunter_matplotlib_2007}, {\sc NumPy} \citep{van_der_walt_numpy_2011}, {\sc scipy} \citep{jones_scipy_2001} and {\sc Python} from \url{python.org}


\end{acknowledgements}

\bibliographystyle{aa}
\bibliography{references}

\appendix
\onecolumn
\section{Catalogue of Galactic LMXBs}
\tiny

\begin{longtable}{llllllllll}\caption{Catalogue of Galactic LMXBs: general information. Confirmed systems are indicated in bold. "Compact object" provides the nature of the accreting compact object; "Companion spec. type" refers to the spectral type of the donor star in the binaries; Right Ascension (RA) and Declination (Dec) are given in J2000 alongside the 90\%\,positional error radius; Gaia Distance is queried from \cite{2021AJ....161..147B}; "Type I burst" indicates the presence of type I X-ray bursts; "Globular cluster" gives the host globular cluster in case of association. $\dagger$/$\dagger\dagger$: reliability flag (see Section\,\ref{subsect:contents}). }\label{tab:cat:general}\\
\hline\hline\\[-2ex]
Identifier &Compact &Companion &R.A. &Dec. &Gaia Dist. &Other Dist. &Type I &Globular \\
&object &spec. type &[deg] &[deg] &[kpc] &[kpc] &burst&cluster \\
\hline\\[-2ex]
\endfirsthead
\caption{continued.}\\
\hline\hline\\[-2ex]
Identifier &Compact &Companion &R.A. &Dec. &Gaia Dist. &Other Dist. &Type I &Globular \\
&object &spec. type &[deg] &[deg] &[kpc] &[kpc] &burst&cluster \\
\hline\\[-2ex]
\endhead
\hline
\endfoot
\bf{NGC 104 125} & NS~$^{\dagger[1]}$ &    & 5.9749 & -72.0639 &  & $4.52(3)$~$^{[2]}$ &  & NGC104\\[3pt]
47 Tuc X-5 & NS~$^{\dagger[3]}$ &    & 6.00413 & -72.08144 &  & $4.52(3)$~$^{[2]}$ &  & NGC104\\[3pt]
47 Tuc X-7 & NS~$^{\dagger[3]}$ &    & 6.01468 & -72.08112 &  & $4.52(3)$~$^{[2]}$ &  & NGC104\\[3pt]
\bf{47 Tuc X-9} & BH~$^{\dagger\dagger[4]}$ &    & 6.01777 & -72.0828 &  & $4.52(3)$~$^{[2]}$ &  & NGC104\\[3pt]
\bf{NGC 104 37} & NS~$^{\dagger\dagger[1]}$ &    & 6.02042 & -72.08083 &  & $4.52(3)$~$^{[2]}$ &  & NGC104\\[3pt]
\bf{NGC 104 17} & NS~$^{\dagger[1]}$ &    & 6.03437 & -72.07555 & 5(2)$^{}$ & $4.52(3)$~$^{[2]}$ &  & NGC104\\[3pt]
\bf{IGR J00291+5934} & NS~$^{[5]}$ &    & 7.26272 & 59.57194 &  & $2.8(8)$~$^{\dagger\dagger[6]}$ &  & \\[3pt]
\bf{3U 0042+32} &    & G0-8~$^{[7]}$ & 11.21 & 33.02139 &  &    &  & \\
\hline
\end{longtable}
\tablebib{[1]\cite{2005ApJ...622..556H}; [2]\cite{2021MNRAS.505.5957B}; [3]\cite{2003ApJ...588..452H}; [4]\cite{2017MNRAS.467.2199B}; [5]\cite{2004ATel..353....1M}; [6]\cite{2008ApJ...672.1079T}; [7]\cite{1978MNRAS.183P..29C}; [8]\cite{2020ApJ...904...49M}; [9]\cite{1993AIPC..280..319K}; [10]\cite{2000MNRAS.317..528W}; [11]\cite{2020MNRAS.494.3912K}; [12]\cite{2016ApJ...831...89S}; [13]\cite{2021ApJ...909L..27Y}; [14]\cite{2022AA...657A.104C}; [15]\cite{2008ApJS..179..360G}; [16]\cite{2015ApJ...798..117P}; [17]\cite{2021ApJ...917...69S}; [18]\cite{2011ATel.3116....1R}; [19]\cite{1995ApJ...440..841B}; [20]\cite{1987ATsir1477....1K}; [21]\cite{1980MNRAS.192..709M}; [22]\cite{2019ATel13270....1T}; [23]\cite{2022MNRAS.515.3105S}; [24]\cite{1985ESASP.236..119P}; [25]\cite{1999IAUC.7174....1G}; [26]\cite{1992PASJ...44..641A}; [27]\cite{2020ATel13452....1K}; [28]\cite{2021AA...650A..69C}; [29]\cite{2017AA...598A..34S}; [30]\cite{2001ApJ...553..335J}; [31]\cite{2005AA...441..675I}; [32]\cite{2007ApJ...669L..85S}; [33]\cite{1999AA...344..101S}; [34]\cite{2002IAUC.7893....2R}; [35]\cite{1999PASP..111..969F}; [36]\cite{2009Sci...324.1411A}; [37]\cite{2022MNRAS.513...71S}; [38]\cite{2020MNRAS.498..728B}; [39]\cite{2019AA...622A.211C}; [40]\cite{2000PASJ...52L..15U}; [41]\cite{2006ApJ...644L..49G}; [42]\cite{2016ApJ...825...46W}; [43]\cite{2015ApJ...806...92W}; [44]\cite{2014ATel.5890....1R}; [45]\cite{2014MNRAS.444.3004D}; [46]\cite{2006ATel..837....1K}; [47]\cite{1984IAUC.3932....2C}; [48]\cite{2013PASJ...65L..10M}; [49]\cite{2021MNRAS.506..581M}; [50]\cite{2002AA...392..885C}; [51]\cite{2002AA...392..931C}; [52]\cite{1985IAUC.4044....1V}; [53]\cite{2019ATel12447....1S}; [54]\cite{2021AA...647A...7L}; [55]\cite{2013MNRAS.434.2696S}; [56]\cite{2004ApJ...613L.133C}; [57]\cite{2015ApJ...804L..12S}; [58]\cite{2014ATel.5765....1B}; [59]\cite{2015PASJ...67...30S}; [60]\cite{1999AA...351L..33O}; [61]\cite{2009MNRAS.394.1597K}; [62]\cite{2018MNRAS.478.4317H}; [63]\cite{2002MNRAS.334..233T}; [64]\cite{2024MNRAS.527.5949Y}; [65]\cite{2017ATel10708....1N}; [66]\cite{2019MNRAS.487.4221S}; [67]\cite{2011ApJ...735..104K}; [68]\cite{2011ATel.3341....1M}; [69]\cite{2015ApJ...803L..27B}; [70]\cite{2017ApJ...849...21B}; [71]\cite{1998ApJ...499..375O}; [72]\cite{2003AA...397..249S}; [73]\cite{2003ApJ...599..498J}; [74]\cite{2004ApJ...616L.139W}; [75]\cite{2002ApJ...568..845O}; [76]\cite{1998IAUC.6842....2M}; [77]\cite{2003MNRAS.346..684J}; [78]\cite{2002ApJ...568..901W}; [79]\cite{2014PASJ...66....6K}; [80]\cite{2007AA...470..331M}; [81]\cite{2010MNRAS.408.1866R}; [82]\cite{2021MNRAS.508.1389C}; [83]\cite{2015MNRAS.449L...1M}; [84]\cite{2017ATel11067....1B}; [85]\cite{2018ATel11272....1C}; [86]\cite{2012ApJ...750L...3K}; [87]\cite{2020AAS...23517009N}; [88]\cite{2020ApJ...893...30X}; [89]\cite{2023ApJ...944...68R}; [90]\cite{1980MNRAS.190P..33P}; [91]\cite{2021ApJ...920..142H}; [92]\cite{1986ApJ...304..664P}; [93]\cite{2001AA...375..447A}; [94]\cite{2023MNRAS.524L..15K}; [95]\cite{2004NuPhS.132..542P}; [96]\cite{2010AA...516A..94N}; [97]\cite{1977ApJ...217L..23H}; [98]\cite{2003AA...403L..11G}; [99]\cite{1972NPhS..238...71G}; [100]\cite{2001IAUC.7707....2M}; [101]\cite{2004ApJ...616..376O}; [102]\cite{2009ATel.2107....1M}; [103]\cite{1995Natur.378..157B}; [104]\cite{1972ApJ...174L.143T}; [105]\cite{1994ApJ...436..319A}; [106]\cite{2023MNRAS.522.2065M}; [107]\cite{2018ATel11342....1B}; [108]\cite{2010ATel.2881....1K}; [109]\cite{2018MNRAS.475.1036C}; [110]\cite{2018AA...610L...2S}; [111]\cite{2013AstL...39..523R}; [112]\cite{2006ATel..748....1H}; [113]\cite{2009ApJ...699...60L}; [114]\cite{2008ATel.1616....1M}; [115]\cite{2003ApJ...583L..95H}; [116]\cite{2017ApJ...846..132H}; [117]\cite{1998AA...332..845K}; [118]\cite{2017AA...608A..31N}; [119]\cite{2012ATel.4219....1D}}
\tablebib{[120]\cite{1983ApJ...267..291G}; [121]\cite{2002AA...382..104M}; [122]\cite{1997PASP..109..461F}; [123]\cite{2004NuPhS.132..486I}; [124]\cite{1985IAUC.4111....1S}; [125]\cite{2005AA...430..997L}; [126]\cite{1998ApJ...508L.163C}; [127]\cite{2007ESASP.622..373G}; [128]\cite{2001AAS...199.2704M}; [129]\cite{2002AA...390..597I}; [130]\cite{2003MNRAS.342..909M}; [131]\cite{1999IAUC.7247....3C}; [132]\cite{2020ApJ...903...37L}; [133]\cite{2019ATel13247....1C}; [134]\cite{2021MNRAS.503..472O}; [135]\cite{2009MNRAS.392..665G}; [136]\cite{2004ATel..233....1T}; [137]\cite{2011ApJ...738..183P}; [138]\cite{2000AA...358L..71K}; [139]\cite{2005AA...440..287I}; [140]\cite{1976Natur.262..474M}; [141]\cite{2007MNRAS.377.1295J}; [142]\cite{2016ApJS..222...15T}; [143]\cite{2007ATel.1065....1K}; [144]\cite{2007ATel.1069....1M}; [145]\cite{2011MNRAS.417..659A}; [146]\cite{1996AA...314..123M}; [147]\cite{2023MNRAS.526.5209C}; [148]\cite{2012AA...538A.123M}; [149]\cite{2003AAS...20312501S}; [150]\cite{2006AA...455..639C}; [151]\cite{2006smqw.confE..14C}; [152]\cite{2013ApJ...776...20C}; [153]\cite{1999IAUC.7133....3M}; [154]\cite{2006ATel..778....1B}; [155]\cite{2007AA...469L..27C}; [156]\cite{2005ATel..632....1K}; [157]\cite{2018ATel11696....1N}; [158]\cite{2022MNRAS.514.5320W}; [159]\cite{2021MNRAS.507..330S}; [160]\cite{2011MNRAS.415.2373S}; [161]\cite{2023MNRAS.519..519S}; [162]\cite{1989AA...225...79H}; [163]\cite{2003AA...398.1103R}; [164]\cite{1997ApJ...487L..77S}; [165]\cite{2003ApJ...590..999G}; [166]\cite{2018AA...613A..22B}; [167]\cite{1980ApJ...239L..57K}; [168]\cite{1977ApJ...211..866D}; [169]\cite{2001MNRAS.321..776F}; [170]\cite{1976ApJ...207L..95L}; [171]\cite{2018ATel11646....1B}; [172]\cite{1996AAS..120C.291V}; [173]\cite{2022ApJ...935..123L}; [174]\cite{2022ApJ...940...81B}; [175]\cite{1989IAUC.4839....1S}; [176]\cite{2002AA...389L..11M}; [177]\cite{2012ApJ...748....5O}; [178]\cite{2015AA...579A..56B}; [179]\cite{2010MNRAS.404.1591D}; [180]\cite{2024MNRAS.527.7603S}; [181]\cite{2013MNRAS.434.1586A}; [182]\cite{2001AstL...27..297P}; [183]\cite{1995IAUC.6207....3W}; [184]\cite{2018ATel11507....1S}; [185]\cite{1988AA...192..147V}; [186]\cite{2005ATel..622....1B}; [187]\cite{2007ApJ...657L..97K}; [188]\cite{1997ApJ...487L..73C}; [189]\cite{2023AA...674A.100D}; [190]\cite{2002AA...389L..43I}; [191]\cite{2000ApJ...533L.131B}; [192]\cite{1997AA...323..158M}; [193]\cite{1996AA...314L..21G}; [194]\cite{2013AA...556A..30B}; [195]\cite{2012MNRAS.422.2661S}; [196]\cite{2015MNRAS.448.1900W}; [197]\cite{2014MNRAS.441..640B}; [198]\cite{2017AstL...43..656M}; [199]\cite{2020MNRAS.492.4344S}; [200]\cite{2017ATel10256....1M}; [201]\cite{1996Natur.381..291F}; [202]\cite{2007MNRAS.380.1511G}; [203]\cite{2014ApJ...792..109D}; [204]\cite{2006ESASP.604..259D}; [205]\cite{2016AA...589A..42B}; [206]\cite{2015ApJ...808...34P}; [207]\cite{1999AA...346L..45C}; [208]\cite{2021ApJ...918....9P}; [209]\cite{2013MNRAS.432.1133M}; [210]\cite{2020AA...637A...2C}; [211]\cite{1996PASJ...48..417M}; [212]\cite{2005AA...443..571P}; [213]\cite{1976IAUC.2922....1L}; [214]\cite{2005AA...430L...9P}; [215]\cite{2010ApJ...712L..58A}; [216]\cite{1997ApJ...486..355S}; [217]\cite{2021MNRAS.506.5619W}; [218]\cite{2017ApJ...834...88M}; [219]\cite{2012ApJ...745L...7S}; [220]\cite{2006AA...456.1105D}; [221]\cite{2014MNRAS.440..365T}; [222]\cite{2019MNRAS.487.2296T}; [223]\cite{2000ApJ...543L..73N}; [224]\cite{2004AA...416..311W}; [225]\cite{2006ATel..970....1B}; [226]\cite{2007ATel.1207....1D}; [227]\cite{2008ATel.1651....1A}; [228]\cite{2011MNRAS.410..179C}; [229]\cite{1990MNRAS.243...72S}; [230]\cite{2005AdSpR..35.1137M}; [231]\cite{1983ApJ...267..310M}; [232]\cite{2014ApJ...793..128V}; [233]\cite{2023MNRAS.521..433T}; [234]\cite{2010ATel.2929....1S}; [235]\cite{2020ApJS..249...32G}; [236]\cite{2020ApJ...904..147U}; [237]\cite{1998IAUC.6964....2F}; [238]\cite{2009ApJ...693.1775O}; [239]\cite{2000IAUC.7482....2M}; [240]\cite{2012ATel.4264....1A}; [241]\cite{2006ApJ...639L..31B}; [242]\cite{2003ApJ...598..481K}; [243]\cite{1999AA...345..100I}; [244]\cite{2001ApJ...563L..41I}; [245]\cite{1985IAUC.4058....2P}; [246]\cite{2020ATel14124....1N}; [247]\cite{2021ApJ...908L..15N}; [248]\cite{2009MNRAS.393..126W}; [249]\cite{2023MNRAS.524.2477A}; [250]\cite{2007AA...461L..17W}; [251]\cite{2007ApJ...657L.109P}; [252]\cite{2003ATel..169....1H}; [253]\cite{2011ATel.3556....1P}; [254]\cite{1987MNRAS.226...39S}; [255]\cite{2018MNRAS.476..354C}; [256]\cite{2002ApJ...575L..21M}; [257]\cite{1999ApJ...523L..45N}; [258]\cite{2002ApJ...575.1018K}; [259]\cite{2009ATel.2198....1B}; [260]\cite{2009ATel.2269....1S}; [261]\cite{2012MNRAS.423.2656R}; [262]\cite{2010ApJ...723.1817S}; [263]\cite{2001AA...378L..37C}; [264]\cite{2005ATel..550....1M}; [265]\cite{2008ApJ...681.1458Z}; [266]\cite{1999NuPhS..69..228I}; [267]\cite{2007ATel.1094....1C}; [268]\cite{1997IAUC.6746....2P}; [269]\cite{2007ATel.1108....1M}; [270]\cite{2021ASPC..528..391S}; [271]\cite{2022sf2a.conf...39D}; [272]\cite{2018AA...618A.150F}; [273]\cite{1995MNRAS.274L..15P}; [274]\cite{2018ATel11957....1F}; [275]\cite{2020AA...641A..37K}; [276]\cite{2007ATel.1054....1B}; [277]\cite{1991MNRAS.248..751V}; [278]\cite{1991PAZh...17..116S}; [279]\cite{2016AA...596A..46M}; [280]\cite{2017ApJ...834...71V}; [281]\cite{2022ApJ...927..151S}; [282]\cite{2012ATel.4050....1C}; [283]\cite{2016AA...587A.102B}; [284]\cite{2017MNRAS.464..170K}; [285]\cite{1998IAUC.6867....2M}; [286]\cite{2007MNRAS.380.1637C}; [287]\cite{2004AA...423L...9K}; [288]\cite{2021MNRAS.501..261A}; [289]\cite{1998AA...338L..83G}; [290]\cite{2020MNRAS.495..796G}; [291]\cite{2008ATel.1443....1M}; [292]\cite{2015MNRAS.450.2915W}; [293]\cite{2017AstL...43..781C}; [294]\cite{2000ApJ...536..891N}; [295]\cite{2017MNRAS.466..906S}; [296]\cite{2017ATel10567....1I}; [297]\cite{2019MNRAS.486.4149G}; [298]\cite{2022MNRAS.513.6196R}; [299]\cite{2018ATel11356....1R}; [300]\cite{2003ATel..164....1M}; [301]\cite{2017MNRAS.466.2261W}; [302]\cite{2022ApJ...935...36D}; [303]\cite{1995PASJ...47..575M}; [304]\cite{1999MNRAS.306..417B}; [305]\cite{2000AA...357..527C}; [306]\cite{1997ApJ...490L.157W}; [307]\cite{2002AA...382..947K}; [308]\cite{2022ATel15425....1B}; [309]\cite{2023MNRAS.521..881M}; [310]\cite{2017ATel10612....1P}; [311]\cite{2006ATel..714....1R}; [312]\cite{2009AA...501....1C}; [313]\cite{2017ApJ...847...44Q}; [314]\cite{2014ApJ...784....2M}; [315]\cite{2001ApJ...555..489O}; [316]\cite{2015ApJ...814..158P}; [317]\cite{2018ApJ...867L...9T}; [318]\cite{2019ApJ...882L..21T}; [319]\cite{2022ApJ...930....9M}; [320]\cite{1976ApJ...205L.127G}; [321]\cite{1987ApJ...322..842R}; [322]\cite{2013ATel.4959....1P}; [323]\cite{2003ApJ...594..798B}; [324]\cite{2007MNRAS.376.1886S}; [325]\cite{2001ApJ...553L..43J}; [326]\cite{1980ApJ...242L.109M}; [327]\cite{2014AA...563A.124F}; [328]\cite{2019PASJ...71..108O}; [329]\cite{1999ApJ...514L..27U}; [330]\cite{1991ApJ...370L..77K}; [331]\cite{2022ApJ...928L...8D}; [332]\cite{2011ATel.3628....1M}; [333]\cite{2019MNRAS.482.2149L}; [334]\cite{2014MNRAS.439.1381R}; [335]\cite{1998AA...329L..37I}; [336]\cite{2021MNRAS.506.4107P}; [337]\cite{1976IAUC.2963....1S}; [338]\cite{2013MNRAS.432.1361C}; [339]\cite{2008ATel.1716....1M}; [340]\cite{2011ATel.3121....1S}; [341]\cite{1994IAUC.6096....1Z}; [342]\cite{2022MNRAS.517.2801W}; [343]\cite{2022ApJ...927..190P}; [344]\cite{2018ApJ...865...33H}; [345]\cite{2012ApJ...759....8D}; [346]\cite{2014ATel.5972....1I}; [347]\cite{2001ApJ...550L.155H}; [348]\cite{2015ATel.7233....1S}; [349]\cite{2020MNRAS.499..793B}; [350]\cite{2021MNRAS.503.5600B}; [351]\cite{1999AAS...19512602F}; [352]\cite{2022MNRAS.517.1476Y}; [353]\cite{2005ATel..523....1M}; [354]\cite{2005AAS...207.3204S}; [355]\cite{2010AA...510A..61T}; [356]\cite{1985AA...147L...3C}; [357]\cite{2007ApJ...665L.147J}; [358]\cite{2004ApJ...609..977G}; [359]\cite{2006MNRAS.365.1387C}; [360]\cite{2013ApJ...778..155R}; [361]\cite{2016cosp...41E2124Y}; [362]\cite{2023AdSpR..71.1045N}; [363]\cite{2013ApJ...768..185S}; [364]\cite{2001ApJ...549L..85G}; [365]\cite{2013MNRAS.431L..10R}; [366]\cite{2020AA...633A..99C}; [367]\cite{1983PASJ...35..531M}; [368]\cite{2022MNRAS.516L..76S}; [369]\cite{2017ApJ...851..114R}; [370]\cite{2021RAA....21..214S}; [371]\cite{1995MNRAS.277L..45C}; [372]\cite{1996PASP..108..762H}; [373]\cite{2002AA...384..163C}; [374]\cite{1998IAUC.6931....1C}; [375]\cite{1994MNRAS.271L...5C}; [376]\cite{1993MNRAS.265..834C}; [377]\cite{2021MNRAS.507.3423M}; [378]\cite{2012ApJ...761....4O}; [379]\cite{2002MNRAS.329...29C}; [380]\cite{2015ApJ...807...52A}; [381]\cite{2022ATel15640....1H}; [382]\cite{1990Natur.347..534D}; [383]\cite{1987ApJ...313L..59G}; [384]\cite{1989ApJ...341L..75G}; [385]\cite{2020MNRAS.495.4508E}; [386]\cite{1998ApJ...493L..39C}; [387]\cite{2014ApJ...787...67D}; }\clearpage
\begin{landscape}
\begin{longtable}{lllllllll}\caption{Catalogue of Galactic LMXBs: orbital data. Confirmed systems are indicated in bold. $Mx$ and $Mo$ refer respectively to the mass of the compact object and the companion star;$K2$ is the semi-amplitude of the radial velocity curve of the companion star; $RV$ is the systemic radial velocity. $\dagger$/$\dagger\dagger$: reliability flag (see Section\,\ref{subsect:contents}). }\label{tab:cat:orbital}\\
\hline\hline\\[-2ex]
Identifier &Mx [\Msun\,] &Mo [\Msun\,] &Period [hr] &Eccentricity &K2 [km/s]&RV [km/s]&Spin period [s] &$\dot{P}/P$ [1/yr] \\
\hline\\[-2ex]
\endfirsthead
\caption{continued.}\\
\hline\hline\\[-2ex]
Identifier &Mx [\Msun\,] &Mo [\Msun\,] &Orbital period [hr] &Eccentricity &K2 [km/s]&RV [km/s]&Spin period [s] &$\dot{P}/P$ [1/yr] \\
\hline\\[-2ex]
\endhead
\hline
\endfoot
\bf{NGC 104 125} &    &    &    &    &    &    &    &   \\[3pt]
47 Tuc X-5 & 1.4~$^{\dagger[1]}$ &    & $8.670(8)$~$^{[1]}$ &    &    &    &    &   \\[3pt]
47 Tuc X-7 & 1.8~$^{\dagger[1]}$ &    & 5.5~$^{\dagger[1]}$ &    &    &    &    &   \\[3pt]
\bf{47 Tuc X-9} &    &    & $0.4696(3)$~$^{[2]}$ &    &    &    &    &   \\[3pt]
\bf{NGC 104 37} &    &    & $3.087(4)$~$^{[3]}$ &    &    &    &    &   \\[3pt]
\bf{NGC 104 17} &    &    &    &    &    &    &    &   \\[3pt]
\bf{IGR J00291+5934} &    & $0.07(4)$~$^{\dagger[4]}$ & $2.456692(2)$~$^{[5]}$ & $<$2.0e-4~$^{\dagger[5]}$ &    &    & $0.00166974977400(3)$~$^{[5]}$ & $<$-4.2e-8~$^{\dagger[5]}$\\[3pt]
\bf{3U 0042+32} &    &    & $2.8(2)e+02$~$^{[6]}$ &    &    &    &    &   \\
\hline
\end{longtable}
\end{landscape}
\tablebib{[1]\cite{2003ApJ...588..452H}; [2]\cite{2017MNRAS.467.2199B}; [3]\cite{2005ApJ...622..556H}; [4]\cite{2008ApJ...672.1079T}; [5]\cite{2005ApJ...622L..45G}; [6]\cite{1978MNRAS.183P..35W}; [7]\cite{2022MNRAS.516.2023C}; [8]\cite{1995ApJ...455..614F}; [9]\cite{2020MNRAS.494.3912K}; [10]\cite{2016ApJ...831...89S}; [11]\cite{2021ApJ...920..120Y}; [12]\cite{2021ApJ...920..121Y}; [13]\cite{2015ApJ...798..117P}; [14]\cite{2011MNRAS.414L..41F}; [15]\cite{2021ApJ...917...69S}; [16]\cite{2012MNRAS.420.3538C}; [17]\cite{2014AA...572A..99B}; [18]\cite{2008ApJ...672L..37S}; [19]\cite{2017MNRAS.472.1907V}; [20]\cite{1986ApJ...308..110M}; [21]\cite{2008MNRAS.384..849N}; [22]\cite{1994MNRAS.266..137M}; [23]\cite{2022MNRAS.515.3105S}; [24]\cite{1997ApJ...486.1000H}; [25]\cite{2009MNRAS.399.2055B}; [26]\cite{2012MNRAS.420...75R}; [27]\cite{2009ATel.2094....1G}; [28]\cite{2020ATel13452....1K}; [29]\cite{2017AA...598A..34S}; [30]\cite{2005AA...441..675I}; [31]\cite{2011ApJ...729....8Z}; [32]\cite{2007ApJ...669L..85S}; [33]\cite{2012MNRAS.424..620A}; [34]\cite{2005MNRAS.356..621J}; [35]\cite{2002ApJ...576L.137G}; [36]\cite{1999PASP..111..969F}; [37]\cite{2020MNRAS.498..728B}; [38]\cite{2009Sci...324.1411A}; [39]\cite{2005AJ....130..759T}; [40]\cite{2012ApJ...744L..25G}; [41]\cite{2008ApJ...679..732G}; [42]\cite{2016ApJ...825...46W}; [43]\cite{2015ApJ...806...92W}; [44]\cite{2015MNRAS.454.2190D}; [45]\cite{2014MNRAS.444.3004D}; [46]\cite{2014ATel.5890....1R}; [47]\cite{2007MNRAS.380.1182B}; [48]\cite{1987ApJ...313..792M}; [49]\cite{2021MNRAS.506..581M}; [50]\cite{2016AA...589A..34G}; [51]\cite{2023AA...669A..57T}; [52]\cite{2022MNRAS.517L..21C}; [53]\cite{2019ATel12434....1K}; [54]\cite{2016ApJ...822...99C}; [55]\cite{2015MNRAS.454.2199M}; [56]\cite{2015MNRAS.450.4292T}; [57]\cite{2004ApJ...613L.133C}; [58]\cite{2015ApJ...804L..12S}; [59]\cite{2016ApJ...820....6C}; [60]\cite{1999AA...351L..33O}; [61]\cite{2018MNRAS.478.4317H}; [62]\cite{2002MNRAS.334..233T}; [63]\cite{2024MNRAS.527.5949Y}; [64]\cite{2019MNRAS.487.4221S}; [65]\cite{2007ATel.1209....1K}; [66]\cite{2016ApJ...827...88C}; [67]\cite{2020AdSpR..65..693M}; [68]\cite{2017ApJ...849...21B}; [69]\cite{1998ApJ...499..375O}; [70]\cite{2004ApJ...616L.139W}; [71]\cite{2002ApJ...568..845O}; [72]\cite{1991PASP..103..636S}; [73]\cite{2008ApJ...685..428M}; [74]\cite{2020MNRAS.494.4382S}; [75]\cite{2002ApJ...568..901W}; [76]\cite{2001ApJ...553L.157M}; [77]\cite{2010MNRAS.408.1866R}; [78]\cite{2023AA...676L...2L}; [79]\cite{2021MNRAS.508.1389C}; [80]\cite{2015MNRAS.449L...1M}; [81]\cite{2016MNRAS.459.3596H}; [82]\cite{2002ApJ...568..273S}; [83]\cite{2019ApJ...884..168G}; [84]\cite{2012ApJ...750L...3K}; [85]\cite{2001ApJ...550..962S}; [86]\cite{2015PKAS...30..593L}; [87]\cite{2023ApJ...944...68R}; [88]\cite{1980MNRAS.190P..33P}; [89]\cite{2021ApJ...920..142H}; [90]\cite{2007JApA...28..175J}; [91]\cite{2019ApJ...878..121I}; [92]\cite{2014ApJ...789...57S}; [93]\cite{2007ApJ...657..994P}; [94]\cite{2018MNRAS.476..354C}; [95]\cite{1990AA...234..181V}; [96]\cite{1997AAS...190.3603Z}; [97]\cite{2004ApJ...616..376O}; [98]\cite{2014MNRAS.437.2554M}; [99]\cite{1995Natur.378..157B}; [100]\cite{1995Natur.375..464H}; [101]\cite{1997MNRAS.288...43R}; [102]\cite{1981ApJ...247.1003D}; [103]\cite{2009AA...500..883S}; [104]\cite{1972ApJ...174L.143T}; [105]\cite{2023MNRAS.522.2065M}; [106]\cite{2018ATel11311....1K}; [107]\cite{2021MNRAS.501.2174T}; [108]\cite{2011arXiv1102.2102K}; [109]\cite{2018AA...610L...2S}; [110]\cite{2018MNRAS.473.3490I}; [111]\cite{2018MNRAS.481L..94P}; [112]\cite{2001ApJ...549L..71W}; [113]\cite{2017ApJ...846..132H}; [114]\cite{2009AA...505.1143I}; [115]\cite{1997ApJ...490..401W}; [116]\cite{2017AA...608A..31N}; [117]\cite{1999ApJ...512L.125M}; [118]\cite{2018ApJ...858L..13S}; [119]\cite{2021ApJ...912..120B}; [120]\cite{2002ApJ...580.1065G}; [121]\cite{1997PASP..109..461F}; [122]\cite{2015ApJ...807..108I}; [123]\cite{2012MNRAS.422L..91W}; [124]\cite{2011MNRAS.413....2J}; [125]\cite{2009AA...506..857I}; [126]\cite{2007MNRAS.377.1295J}; [127]\cite{2012NewA...17...43G}; [128]\cite{2012ApJ...746....9P}; [129]\cite{2023MNRAS.526.5209C}; [130]\cite{1996AA...314..123M}; [131]\cite{2013ApJ...776...20C}; [132]\cite{2021MNRAS.508.3812N}; [133]\cite{2022MNRAS.514.5320W}; [134]\cite{2021MNRAS.507..330S}; [135]\cite{2023MNRAS.519..519S}; [136]\cite{1988ApJ...331..764H}; [137]\cite{2003ApJ...593L..35S}; [138]\cite{2020MNRAS.495L..37V}; [139]\cite{1997ApJ...487L..77S}; [140]\cite{2018AA...613A..22B}; [141]\cite{2006ApJ...641..479H}; [142]\cite{2016pas..conf..133M}; [143]\cite{1990IAUC.5104....1S}; [144]\cite{2018MNRAS.478..448J}; [145]\cite{2001MNRAS.321..776F}; [146]\cite{2022ApJ...935..123L}; [147]\cite{2012ApJ...748....5O}; [148]\cite{2003AA...399..699R}; [149]\cite{1997ApJ...479L.137S}; [150]\cite{2024MNRAS.527.7603S}; [151]\cite{1995IAUC.6207....3W}; [152]\cite{2019ApJ...877...70B}; [153]\cite{2018ATel11507....1S}; [154]\cite{1986MNRAS.222P..15C}; [155]\cite{2006MNRAS.373.1235C}; [156]\cite{2023MNRAS.521.5904B}; [157]\cite{2018PASJ...70...67W}; [158]\cite{1996Natur.381..291F}; [159]\cite{2014ATel.5901....1P}; [160]\cite{2021ATel14695....1N}; [161]\cite{2015MNRAS.447.3034H}; [162]\cite{2005MNRAS.357.1211S}; [163]\cite{2016AA...595A..52Z}; [164]\cite{2016AA...589A..42B}; [165]\cite{2020AA...637A...2C}; [166]\cite{2013MNRAS.432.1133M}; [167]\cite{2022MNRAS.514.5375B}; [168]\cite{2017MNRAS.464..840P}; [169]\cite{2005AA...443..571P}; [170]\cite{1997ApJ...486..355S}; [171]\cite{2005AA...430L...9P}; [172]\cite{2010ApJ...712L..58A}; [173]\cite{2017ApJ...834...88M}; [174]\cite{2010MNRAS.401.1255J}; [175]\cite{2019MNRAS.487.2296T}; [176]\cite{2014ApJ...793..128V}; [177]\cite{2009ApJ...693.1775O}; [178]\cite{2011AA...526L...3P}; [179]\cite{2006ApJ...639L..31B}; [180]\cite{2013ApJ...765L...1G}; [181]\cite{2017ApJ...844...53C}; [182]\cite{2008ApJ...674L..45A}; [183]\cite{2020ATel14124....1N}; [184]\cite{2021ApJ...908L..15N}; [185]\cite{2022MNRAS.515.3838M}; [186]\cite{2011ApJ...727L..18A}; [187]\cite{2011AA...535L...4P}; [188]\cite{2015ApJ...798...56L}; [189]\cite{2004MNRAS.347..334B}; [190]\cite{2009AA...508..297D}; [191]\cite{2002ApJ...575L..21M}; [192]\cite{2002ApJ...575.1018K}; [193]\cite{2010MNRAS.407.2575P}; [194]\cite{2003AA...406..233I}; [195]\cite{2010ApJ...723.1817S}; [196]\cite{2012MNRAS.423.2656R}; [197]\cite{2008ApJ...681.1458Z}; [198]\cite{2017MNRAS.471.2508S}; [199]\cite{2007ApJ...668L.147K}; [200]\cite{2018ApJ...864...14B}; [201]\cite{2021ASPC..528..391S}; [202]\cite{2005ApJ...618L..45P}; [203]\cite{1984ApJ...283L...9W}; [204]\cite{2020ApJ...899..120T}; [205]\cite{1998ApJ...504L..35W}; [206]\cite{2006PhDT........26B}; [207]\cite{2002ApJ...578L.129S}; [208]\cite{2021ATel14606....1H}; [209]\cite{2016AA...587A.102B}; [210]\cite{2007MNRAS.380.1637C}; [211]\cite{2011ApJ...742...17L}; [212]\cite{2003IAUC.8095....2M}; [213]\cite{2004AA...423L...9K}; [214]\cite{2009MNRAS.395..884E}; [215]\cite{2008MNRAS.391.1619D}; [216]\cite{1998Natur.394..346C}; [217]\cite{2008int..workE..78B}; [218]\cite{2017MNRAS.466..906S}; [219]\cite{2020AA...639A..33S}; [220]\cite{2017MNRAS.466.2261W}; [221]\cite{2007MNRAS.375..971P}; [222]\cite{2003ATel..164....1M}; [223]\cite{2022ApJ...935...36D}; [224]\cite{2003ApJ...595.1086C}; [225]\cite{2022ApJ...931L...9A}; [226]\cite{1986ASIC..167..253L}; [227]\cite{1997ApJ...490L.157W}; [228]\cite{2022ApJ...935L..32B}; [229]\cite{2021ApJ...912..110B}; [230]\cite{2017ApJ...847...44Q}; [231]\cite{2014ApJ...784....2M}; [232]\cite{2001ApJ...555..489O}; [233]\cite{2005MNRAS.363..882L}; [234]\cite{2020ApJ...893L..37T}; [235]\cite{2019ApJ...882L..21T}; [236]\cite{2011MNRAS.415.3344K}; [237]\cite{1987ApJ...322..842R}; [238]\cite{2023ApJ...951...42C}; [239]\cite{2013Natur.501..517P}; [240]\cite{2007MNRAS.376.1886S}; [241]\cite{2015AA...577A..63I}; [242]\cite{2010MNRAS.409..755J}; [243]\cite{2003MNRAS.339..663J}; [244]\cite{2001ApJ...553L..43J}; [245]\cite{2010AstL...36..738M}; [246]\cite{2022ApJ...928L...8D}; [247]\cite{1991ApJ...370L..77K}; [248]\cite{2014MNRAS.439.1381R}; [249]\cite{2012ApJ...747..119E}; [250]\cite{2000ApJ...530L..21D}; [251]\cite{2013MNRAS.432.1361C}; [252]\cite{2011ATel.3121....1S}; [253]\cite{2020AAS...23515902S}; [254]\cite{2001ApJ...550L.155H}; [255]\cite{1986ApSS.126...89P}; [256]\cite{2022MNRAS.514.1908K}; [257]\cite{2021MNRAS.503.5600B}; [258]\cite{2022MNRAS.517.1476Y}; [259]\cite{2006ApJ...638..963K}; [260]\cite{2017ApJ...840....9M}; [261]\cite{2023AdSpR..71.1045N}; [262]\cite{2012ATel.4347....1C}; [263]\cite{2000AJ....120..943W}; [264]\cite{2008ApJ...674L..41C}; [265]\cite{2013ApJ...768..185S}; [266]\cite{1986ApJ...304..231N}; [267]\cite{2001MNRAS.322..827H}; [268]\cite{2001ApJ...549L..85G}; [269]\cite{2013MNRAS.431L..10R}; [270]\cite{2022MNRAS.516L..76S}; [271]\cite{2021RAA....21..214S}; [272]\cite{2011ApJ...730...43B}; [273]\cite{2004AJ....127..481I}; [274]\cite{1995MNRAS.277L..45C}; [275]\cite{1996PASP..108..762H}; [276]\cite{2023ApJ...954...62D}; [277]\cite{1992ApJ...401L..97W}; [278]\cite{1994MNRAS.271L...5C}; [279]\cite{1992Natur.355..614C}; [280]\cite{2002ApJ...581..570T}; [281]\cite{2002MNRAS.329...29C}; [282]\cite{1999ApJ...521..341T}; [283]\cite{2015ApJ...807...52A}; [284]\cite{2005ApJ...634L.105D}; [285]\cite{1993AA...270..139I}; [286]\cite{1986MNRAS.218...63H}; [287]\cite{1979ApJ...233L..57T}; [288]\cite{2020MNRAS.495.4508E}; [289]\cite{2010MNRAS.401.2517C}; [290]\cite{1999MNRAS.305..132O}; }

\newpage
\section{\emph{INTEGRAL} sources}

\begin{table}[h]
\centering
\caption{List of unidentified {\it INTEGRAL} sources from \cite{2016ApJS..223...15B} with available identification in the literature.}\label{tab:IGR}

\begin{tabular}{llllll}
\hline\hline\\[-2ex]
Identifier & R.A. [deg] & Dec. [deg] & Simbad Type & Identification & Ref. \\
\hline\\[-2ex]
  IGR J00486-4241 & 12.151 & -42.719 & gamma & AGN ? & [1]\\
  IGR J04069+5042 & 61.73 & 50.702 & gamma & RS CVn & [2]\\
  IGR J04539+4502 & 73.46 & 45.038 & gamma & AGN & [3]\\
  IGR J05511-1218 & 87.776 & -12.322 & gamma & AGN & [4]\\
  IGR J06552-1146 & 103.792 & -11.77 & gamma & RS CVn ? & [2]\\
  IGR J07072-1227 & 106.801 & -12.427 & gamma & AGN & [4]\\
  IGR J08004-4309 & 120.092 & -43.166 & X & CV & [5]\\
  IGR J08262+4051 & 126.556 & 40.855 & gamma & AGN & [6]\\
  IGR J10252-6829 & 156.25204 & -68.45758 & gamma & early type & [7]\\
  2MASS J10445192-6025115 & 161.21635 & -60.41987 & YSO & AGN & [8]\\
  IGR J12341-6143 & 188.467 & -61.796 & gamma & SFXT & [9]\\
  IGR J12562+2554 & 194.05 & 25.905 & gamma & AGN & [10]\\
  IGR J13045-5630 & 196.13125 & -56.51522 & gamma & AGN & [11]\\
  IGR J13402-6428 & 205.05 & -64.48 & gamma & AGN & [12]\\
  IGR J14043-6148 & 211.123 & -61.789 & gamma & blazar ? & [13]\\
  IGR J14059-6116 & 211.485 & -61.275 & gamma & Gamma HMXB & [14]\\
  IGR J14385+8553 & 219.616 & 85.883 & gamma & AGN & [6]\\
  IGR J14466-3352 & 221.656 & -33.875 & gamma & AGN ? & [15]\\
  IGR J14557-5448 & 223.862 & -54.786 & gamma & AGN ? & [16]\\
  HD 132658 & 224.99804 & 12.02395 & Star & AGN ? & [5]\\
  IGR J15391-5307 & 234.82 & -53.138 & gamma & AGN & [17]\\
  UCAC4 170-140217 & 238.55406 & -56.15832 & Star & AGN ? & [18]\\
  IGR J16173-5023 & 244.314 & -50.386 & gamma & magnetic CV ? & [17]\\
  IGR J16316-4028 & 247.9 & -40.467 & gamma & gamma ray pulsar?  & [19]\\
  IGR J16413-4046 & 250.331 & -40.794 & gamma & AGN ? & [17]\\
  IGR J16447-5138 & 251.178 & -51.649 & gamma & YSO or AGN & [5]\\
  IGR J16460+0849 & 251.488 & 8.818 & gamma & extragalactic & [20]\\
  IGR J16560-4958 & 253.989 & -49.967 & gamma & AGN & [17]\\
  IGR J17259+2603 & 261.521 & 25.905 & gamma & Type 1 QSO & [4]\\
  IGR J17348-2045 & 263.745 & -20.759 & gamma & AGN ? & [21]\\
  IGR J17375-3022 & 264.3946 & -30.3817 & gamma & SFXT & [9]\\
  IGR J18293-1213 & 277.342 & -12.222 & gamma & CV & [22]\\
  SWIFT J1839.1-5717 & 279.7762 & -57.2511 & X & AGN & [5]\\
  IGR J18457+0244 & 281.42 & 2.702 & gamma & AGN & [8]\\
  IGR J18485-0047 & 282.106027 & -0.776356 & gamma & AGN ? & [23]\\
  IGR J18532+0416 & 283.318 & 4.297 & gamma & AGN & [8]\\
  IGR J19386-4653 & 294.653 & -46.886 & gamma & AGN ? & [24]\\
  IGR J20155+3827 & 303.875 & 38.45 & gamma & HMXB & [25]\\
  IGR J21171+3930 & 319.306 & 39.515 & gamma & AGN ? & [3]\\
  IGR J21268+6203 & 321.692 & 62.062 & gamma & AGN ? & [26]\\
  IGR J21485+4306 & 327.127 & 43.104 & gamma & AGN ? & [1]\\
  2MASX J21561518+1722525 & 329.06346 & 17.38128 & X & AGN ? & [27]\\
  IGR J22234-4116 & 335.85 & -41.262 & gamma & AGN ? & [28]\\
  1SWXRT J230642.8+550817 & 346.727 & 55.173 & X & CV & [5]\\
\hline
\end{tabular}

\tablebib{
[1]  \cite{2017ATel10467....1L}; [2]  \cite{2012ApJ...761....4O};
[3]  \cite{2017ATel10468....1M}; [4]  \cite{2017ATel10447....1M};
[5]  \cite{2017MNRAS.470.1107L}; [6]  \cite{2012AA...538A.123M};
[7]  Fortin et al. in prep;      [8]  \cite{2018AA...618A.150F};
[9]  \cite{2020MNRAS.491.4543S}; [10] \cite{2010ApJ...720..987F};
[11] \cite{2019ApJ...881..154P}; [12] \cite{2023ATel16099....1F};
[13] \cite{2013AJ....146..110C}; [14] \cite{2019ApJ...884...93C};
[15] \cite{2010ATel.3065....1L}; [16] \cite{2019ATel12840....1U};
[17] \cite{2012ApJ...754..145T}; [18] \cite{2020ApJ...889...53T};
[19] \cite{2002cxo..prop.1612F}; [20] \cite{2009AA...497..347Z};
[21] \cite{2012AA...544A.118B};  [22] \cite{2016MNRAS.461..304C};
[23] \cite{2009ApJ...701..811T}; [24] \cite{2012AA...548A..32M};
[25] \cite{2021MNRAS.503..472O}; [26] \cite{2011MNRAS.416..531M};
[27] \cite{2016MNRAS.460...19M}; [28] \cite{2010MNRAS.403..945L};
}
\end{table}

\end{document}